\documentclass[english,11pt,aps,prd,a4paper,preprintnumbers,floatfix,nofootinbib,showpacs,superscriptaddress,notitlepage]{revtex4-1} 

\newcommand{\AddrUNAM}{Instituto de F\'isica, Universidad Nacional Aut\'onoma de M\'exico, A.P. 20-364, Ciudad de M\'exico 01000, M\'exico.}

\newcommand{\AddrAHEP}{AHEP Group, Institut de F\'{i}sica Corpuscular --
	C.S.I.C./Universitat de Val\`{e}ncia, Parc Cientific de Paterna.\\
	C/Catedr\'atico Jos\'e Beltr\'an, 2 E-46980 Paterna (Val\`{e}ncia) - SPAIN}

\newcommand{\AddrCinvestav}{Departamento de F\'{\i}sica, Centro de
	Investigaci{\'o}n y de Estudios Avanzados del IPN\\ Apdo. Postal
	14-740 07000 Ciudad de M\'exico, M\'exico}

\usepackage{booktabs}
\usepackage{color}
\usepackage{graphicx}
\usepackage{comment}
\usepackage{multirow}
\usepackage{amsmath}
\usepackage{amssymb}
\usepackage[colorlinks=true,citecolor=darkred,urlcolor=darkred, pdfborder={0 0 0}]{hyperref}
\usepackage[normalem]{ulem}
\definecolor{darkred}{rgb}{0.6,0,0}
\definecolor{drkgrn}{RGB}{0, 51, 0}
\usepackage{soul}

\usepackage{stackengine}
\newcommand\brabar{\scalebox{.3}{(}\raisebox{-1.7pt}{--}\scalebox{.3}{)}}

\newcommand{\eps}{\epsilon}

\begin{document}

\title{Global constraints on neutral-current generalized neutrino interactions}

\author{F. J. Escrihuela}\email{franesfe@alumni.uv.es}\affiliation{\AddrAHEP}
\author{L. J. Flores}\email{luisjf89@fisica.unam.mx}\affiliation{\AddrUNAM}
\author{O. G. Miranda}\email{omr@fis.cinvestav.mx}\affiliation{\AddrCinvestav}
\author{Javier Rend\'{o}n}\email{jrendon@fis.cinvestav.mx}\affiliation{\AddrCinvestav}

\begin{abstract}
{\noindent We study generalized neutrino interactions (GNI) for
  several neutrino processes, including neutrinos from
  electron-positron collisions, neutrino-electron scattering, and
  neutrino deep inelastic scattering. We constrain scalar,
  pseudoscalar, and tensor new physics effective couplings, based on
  the standard model effective field theory at low energies. We have
  performed a global analysis for the different effective
  couplings. We also present the different individual constraints for
  each effective parameter (scalar, pseudoscalar, and tensor). Being a
  global analysis, we show robust results for the restrictions on the
  different GNI parameters and improve some of these bounds.  }
\end{abstract}

\maketitle

\section{Introduction}
Precision neutrino physics experiments have successfully determined
the standard three-neutrino oscillation picture parameters. Future
long-baseline neutrino experiments such as DUNE~\cite{Abi:2020qib} and
Hyper-K~\cite{Abe:2015zbg} will measure this phase accurately.
Current neutrino data allows an exhaustive search on physics beyond
the Standard Model (SM) of Particle Physics. The attempts to explain the
neutrino mass pattern implies models with new interactions that modify
the Standard Model Lagrangian and predict different values for
its vector and axial couplings, a deviation that can be described by
the neutrino non-standard interactions (NSI)
formalism\cite{Ohlsson:2012kf,Miranda:2015dra,Farzan:2017xzy,
	Dev:2019anc}. Besides, electromagnetic neutrino properties could give
a clear signature of new physics if a non-zero neutrino magnetic
moment exists~\cite{Giunti:2014ixa,Miranda:2019wdy}. This tensor
interaction could be tested in the future, for example, in dark
matter direct detection experiments~\cite{AristizabalSierra:2020zod}.
Other kinds of physics beyond the Standard Model, such as leptoquark
models~\cite{Buchmuller:1986zs, Crivellin:2019dwb,
	Gargalionis:2020xvt}, also contain modifications to the usual
couplings and predict new scalar couplings. Moreover, motivated by
experimental data, special attention has been given to scalar neutrino
interactions through solar neutrino measurements from
Borexino~\cite{Ge:2018uhz,Khan:2019jvr}, and even from astrophysical
and cosmological observations like BBN, CMB, and
supernovae~\cite{Heurtier:2016otg,Huang:2017egl,Forastieri:2019cuf,Escudero:2019gvw,Venzor:2020ova}.

From a phenomenological perspective, we can study all these different
interactions in a model-independent framework by considering all the
couplings allowed by the effective field theory. Such a picture is
known as generalized neutrino interactions (GNI) and it can be
considered as an extension of the non-standard interaction
framework. In the GNI case, we consider the appearance of scalar,
pseudoscalar, and tensor couplings, besides the possible modification
to the vector and axial couplings (also considered in NSI). 

Generalized Neutrino Interactions have been recently studied using
measurements from decay branching ratios, $\beta$ decay, and neutrino
scattering off electrons and
quarks~\cite{Bischer:2018zcz,Bischer:2019ttk,Han:2020pff,Li:2020wxi,Chen:2021uuw},
as well as from coherent elastic neutrino-nucleus
scattering~\cite{Lindner:2016wff,Kosmas:2017tsq,
	AristizabalSierra:2018eqm,Li:2020lba}.  
In this work, we
perform a global analysis of different neutrino processes in order to
constrain neutral-current GNI parameters. Besides these global constraints, we will also report several new individual constrains.
In particular, we use
data from neutrino-electron scattering, neutrino Deep Inelastic
Scattering (DIS), and  single-photon detection from
electron-positron collisions.

This work is organized as follows. In Sec.~\ref{sec:formalism}, we
present the general formalism of the generalized neutrino
interactions and the relevant cross-sections for our
analysis. In Sec.~\ref{sec:experimental}, we provide a brief
description of various neutrino experiments used in this work,
their relevant observables, and the $\chi^2$ adopted for each of
them. We report our constraints for each experimental observable in
Sec.~\ref{sec:results}. We also show, in the same section, the global
limits on the GNI parameters. Finally we summarize and give our
conclusions in Sec.~\ref{sec:conclusions}.

\section{The generalized interactions formalism and cross-sections}
\label{sec:formalism}

We introduce in this section the GNI framework that we will use in our
study. The natural approach is to consider an effective field theory
that includes new physics and that, at the electroweak scale, respects
the $SU(3)_\mathrm{C}\times SU(2)_\mathrm{L}\times U(1)_\mathrm{Y}$
gauge group. This approach is usually known as the standard model
effective field theory
(SMEFT)~\cite{Buchmuller:1985jz,Grzadkowski:2010es}. We will adopt
this scheme in our analysis.

We will consider that, additional to the Standard Model Lagrangian, we
can have contributions from the most general effective Lagrangian for
neutral currents that is given by the following four-fermion
interaction~\cite{Bischer:2019ttk,Han:2020pff,Bischer:2018zcz},
\begin{equation}\label{equation_1}
	\mathcal{L}^{NC}_{eff}=-\frac{G_{F}}{\sqrt{2}}\sum_{j}\epsilon^{f,j}_{\alpha\beta}(\bar{\nu}_{\alpha}\mathcal{O}_{j}\nu_{\beta})(\bar{f}\mathcal{O}^{'}_{j}f)\,,\\
\end{equation}
where $G_F$ is the Fermi coupling constant, $f$
represents a fermion with a given flavor, and 
$\nu_{\alpha\beta}$  a (anti)neutrino with flavor
$\alpha,\beta$. The operators $\mathcal{O}_{j}$ and
$\mathcal{O}^{'}_{j}$ account for the generalized interactions that
are explicitly shown in Table~\ref{table_1}. These operators are assumed
to have an effective strength given by the couplings $\epsilon^{f,j}_{\alpha\beta}$
for left-handed neutrino interactions. 
An even more general picture would
include the embedding of right-handed
neutrinos~\cite{delAguila:2008ir,Aparici:2009fh}. For the case of Dirac neutrinos, the right-handed components would be mandatory and may lead to additional constraints if new interactions are considered, such as in the case of the effective number of neutrino species in the early Universe~\cite{Luo:2020sho} or they may even help to disentangle the Dirac vs Majorana nature~\cite{Rodejohann:2017vup}. 
However, this  is out of the
scope of this work.
Several analyses have
set constraints on some of this additional coefficients using different
measurements from LHC, LEP, and the projected sensitivities from
future collider
facilities~\cite{Ruiz:2017nip,Cai:2017mow,Alcaide:2019pnf,Biekotter:2020tbd}.

In defining the Lagrangian in Eq.~(\ref{equation_1}) we have adopted a
convention that is commonly used for these couplings.
For a different convention we refer the reader to the Appendix~A.
It could be
possible to return back to a Lagrangian where the couplings are more
transparent for the reader. For example, we can see that in the case of
a pseudoscalar coupling we have
\begin{equation}
	\bar{\nu}_{\alpha}(1-\gamma^{5})\nu_{\beta}\bar{f}\gamma^{5}f
	= 2 \bar{\nu}_{\alpha R}\nu_{\beta L}\bar{f}\gamma^{5}f
	, \end{equation}
where the pseudoscalar nature of this term is evident in the
right-hand side of the previous equation. Analogous relations can
be found for the rest of the couplings, where the differences restrict
to proportionality constants. An exception is the case of the left and
right-handed couplings ($L,R$) that are linear combinations of
the $V,A$ couplings and coincide with those in the non-standard
interactions (NSI) formalism.

The Lagrangian in Eq.~(\ref{equation_1}) is added to the Standard
Model Lagrangian, that have the left and right handed couplings
\begin{eqnarray}\label{equation_3}
	g^{L,SM}&=&1+\left(-\frac{1}{2}+s^{2}_{W}\right),\nonumber\\
	g^{R,SM}&=&s^{2}_{W},
\end{eqnarray}
where $s_{W}$ is the sine of the weak mixing angle.

\begin{table}[h!]
	\centering
	\begin{tabular}{ccccc}
		\hline
		${\epsilon}$ & & $\mathcal{O}_{j}$ & &$\mathcal{O}^{'}_{j}$\\
		\hline
		$\epsilon^{f,L}$ & &$\gamma_{\mu}(1-\gamma^{5})$ & &$\gamma^{\mu}(1-\gamma^{5})$\\
		\hline
		$\epsilon^{f,R}$ & &$\gamma_{\mu}(1-\gamma^{5})$ & &$\gamma^{\mu}(1+\gamma^{5)}$\\
		\hline
		$\epsilon^{f,S}$ & &$(1-\gamma^{5})$ & & $1$\\
		\hline
		$-\epsilon^{f,P}$ & &$(1-\gamma^{5})$ & &$\gamma^{5}$\\
		\hline
		$\epsilon^{f,T}$ & &$\sigma_{\mu\nu}(1-\gamma^{5})$ & &$\sigma^{\mu\nu}(1-\gamma^{5})$\\
		\hline
	\end{tabular}
	\caption{Effective operators and effective couplings in Eq. (\ref{equation_1}) studied in this work.}
	\label{table_1}
\end{table}

In this work we will compute constraints for the scalar, pseudoscalar,
and tensor interactions. We will focus in left(right) handed
(anti)neutrinos. In the case of vector and axial couplings (or
left and right handed chiralities) there are already several works in
the literature where a current status can be
found~\cite{Ohlsson:2012kf,Miranda:2015dra,Farzan:2017xzy}. For the
sake of completeness, we remind here some of the most relevant
constraints for these couplings.

The most stringent constraint reported in the literature restrict the
muon neutrino couplings, as muon neutrino fluxes are abundant in
accelerator experiments~\footnote{For other  constraints from astrophysical observations, see for example, Refs.~\cite{Mangano:2006ar,Du:2021idh}}. For flavor diagonal couplings, the
constraints for beyond the Standard Model interactions with electrons
are~\cite{Davidson:2003ha,Barranco:2007ej}
\begin{equation}
	|\epsilon^{e,L}_{\mu\mu}| < 3\times10^{-2}, \quad
	|\epsilon^{e,R}_{\mu\mu}| < 3\times10^{-2} ,
\end{equation}
and
\begin{equation}
	|\epsilon^{e,L}_{\mu\tau}| < 0.13, \quad
	|\epsilon^{e,R}_{\mu\tau}| < 0.13.
\end{equation}
For the interactions with quarks, the constraints are given, at $90$ \% CL by~\cite{Escrihuela:2011cf}
\begin{equation}
	|\epsilon^{d,V}_{\mu\mu}| < 0.042, \quad
	|\epsilon^{d,A}_{\mu\mu}| < 0.057,
\end{equation}
while for the flavor changing case we have
\begin{equation}
	|\epsilon^{q,V}_{\mu\tau}| < 7\times10^{-3}, \quad
	|\epsilon^{q,A}_{\mu\tau}| < 39\times10^{-3}.
\end{equation}

In the case of interactions with electrons, the most stringent constraint 
on the right-handed coupling comes from the TEXONO reactor antineutrino
experiment, while for the left-handed coupling a combination of
solar and KamLAND experiments provides the stronger limit~\cite{Barranco:2007ej,Bolanos:2008km,Deniz:2010mp}: 
\begin{equation}
	-0.021 < \epsilon^{e,L}_{ee} < 0.052, \quad
	-0.07 < \epsilon^{e,R}_{ee} < 0.08 ,
\end{equation}
and
\begin{equation}
	| \epsilon^{e,L}_{e\tau} | < 0.33, \quad
	| \epsilon^{e,R}_{e\tau} | < 0.19 .
\end{equation}

For the interactions with quarks, the constraints for electron
neutrinos are given by~\cite{Davidson:2003ha}
\begin{equation}
	|\epsilon^{d,L}_{ee}| < 0.3, \quad
	-0.6 < \epsilon^{d,R}_{ee} < 0.5,
\end{equation}
while for the flavor changing case we have the constraints~\cite{Davidson:2003ha}
\begin{equation}
	| \epsilon^{q,L}_{e\tau} | < 0.5, \quad
	| \epsilon^{q,R}_{e\tau} | < 0.5 .
\end{equation}

Finally, the tau neutrino case have less stringent constraints. For the interaction with electrons they are given by~\cite{Barranco:2007ej,Bolanos:2008km}
\begin{equation}
	-0.12 <  \epsilon^{e,L}_{\tau\tau} < 0.06, \quad
	-0.25 < \epsilon^{e,R}_{\tau\tau} < 0.43,
\end{equation}
and~\cite{Davidson:2003ha}
\begin{equation}
	| \epsilon^{e,L}_{\mu\tau} | < 0.1, \quad
	| \epsilon^{e,R}_{\mu\tau} | < 0.1 .
\end{equation}
For the interaction with quarks, the flavor-diagonal  constraint is~\cite{GonzalezGarcia:2011my}
\begin{equation}
	|\epsilon^{q,V}_{\tau\tau}| < 0.037, 
\end{equation}
while for the flavor-changing case~\cite{Escrihuela:2011cf}
\begin{equation}
	| \epsilon^{q,L}_{\mu\tau} | < 23\times10^{-3}, \quad
	| \epsilon^{q,R}_{\mu\tau} | < 36\times10^{-3}.
\end{equation}

Once we have overviewed the current status for the NSI parameters, we
can turn our attention to the scalar, pseudoscalar, and tensor GNI
parameters. We start by defining the relevant cross-sections for our
study and their modification when a GNI parameter is present. 

\subsection{Cross-section for $e^{+}e^{-}\rightarrow\nu\bar{\nu}\gamma$}
The neutrino-antineutrino pair produced together with a single photon
is a clean signal measured with precision at LEP~\cite{Barate:1997ue,
	Barate:1998ci, Heister:2002ut,Abreu:2000vk,Acciarri:1997dq,
	Acciarri:1998hb, Acciarri:1999kp,Ackerstaff:1997ze, Abbiendi:1998yu,
	Abbiendi:2000hh}. This experimental result has been previously
studied to put constraints on NSI~\cite{Berezhiani:2001rs,Barranco:2007ej,Forero:2011zz} and non-unitarity~\cite{Escrihuela:2019mot}, for instance. 

\begin{figure}[h]
	\begin{minipage}{0.05\textwidth}
		(a)
	\end{minipage}
	\begin{minipage}{0.3\textwidth}
		\includegraphics[scale=0.6]{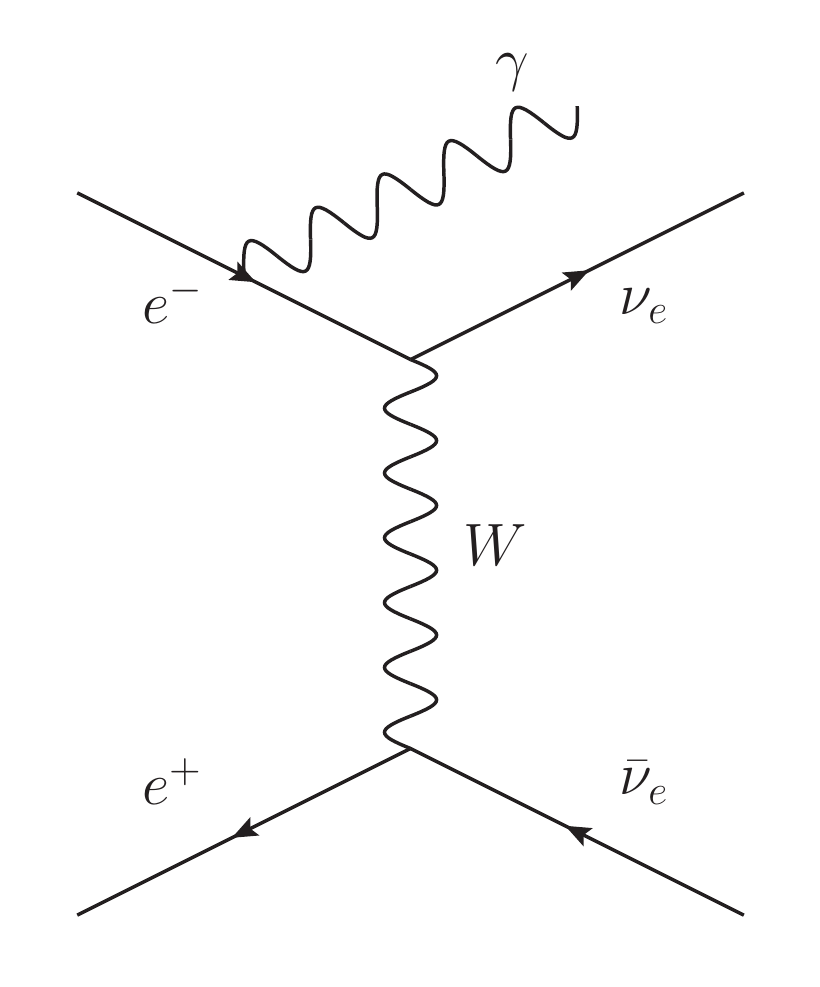}
	\end{minipage}
	\begin{minipage}{0.3\textwidth}\vspace{1.3cm}
		\includegraphics[scale=0.6]{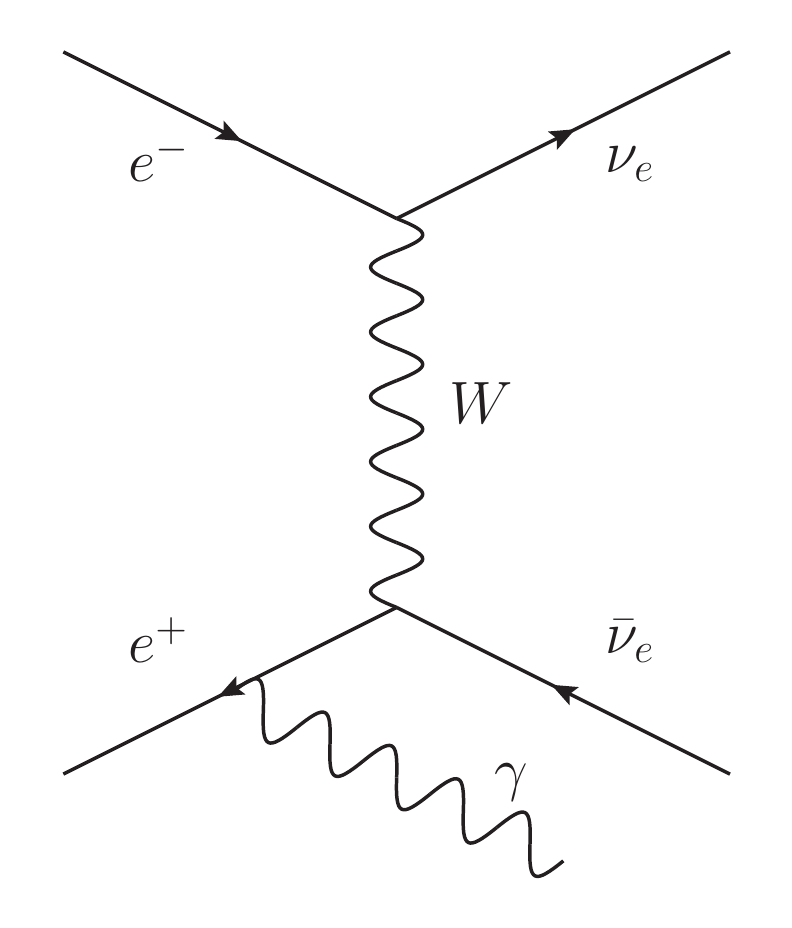}
	\end{minipage}
	\begin{minipage}{0.3\textwidth}
		\includegraphics[scale=0.6]{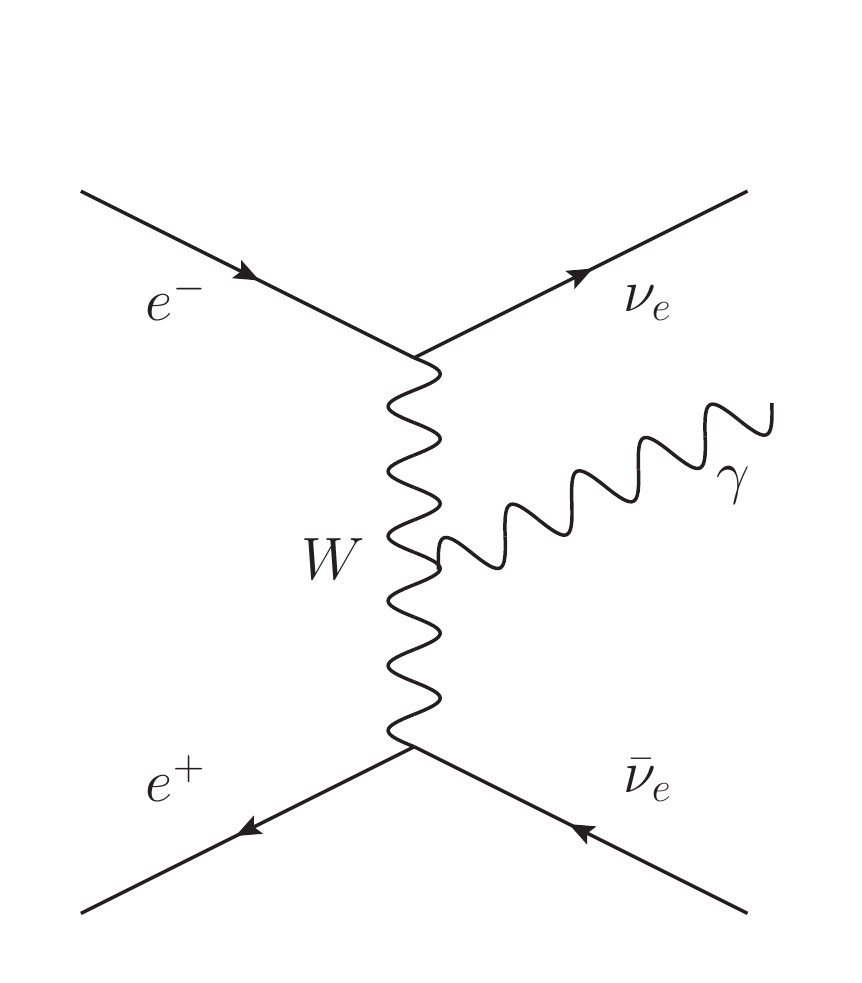}
	\end{minipage}
	\begin{minipage}{0.05\textwidth}
		(b)
	\end{minipage}
	\begin{minipage}{0.4\textwidth}
		\includegraphics[scale=0.6]{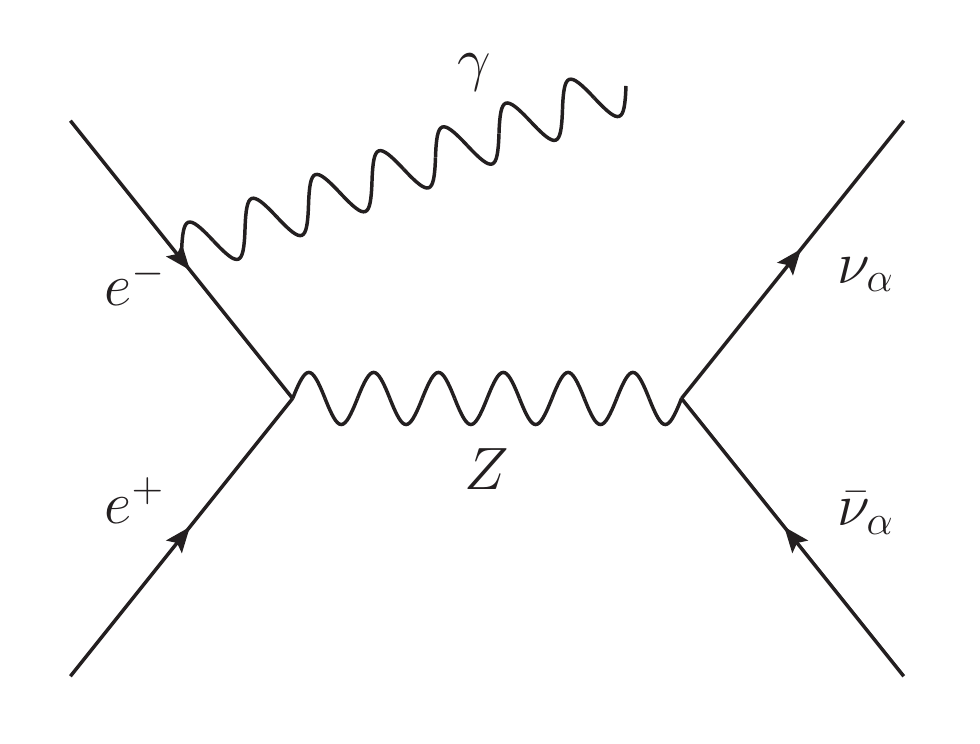}
	\end{minipage}
	\begin{minipage}{0.4\textwidth} \vspace{0.6cm}
		\includegraphics[scale=0.6]{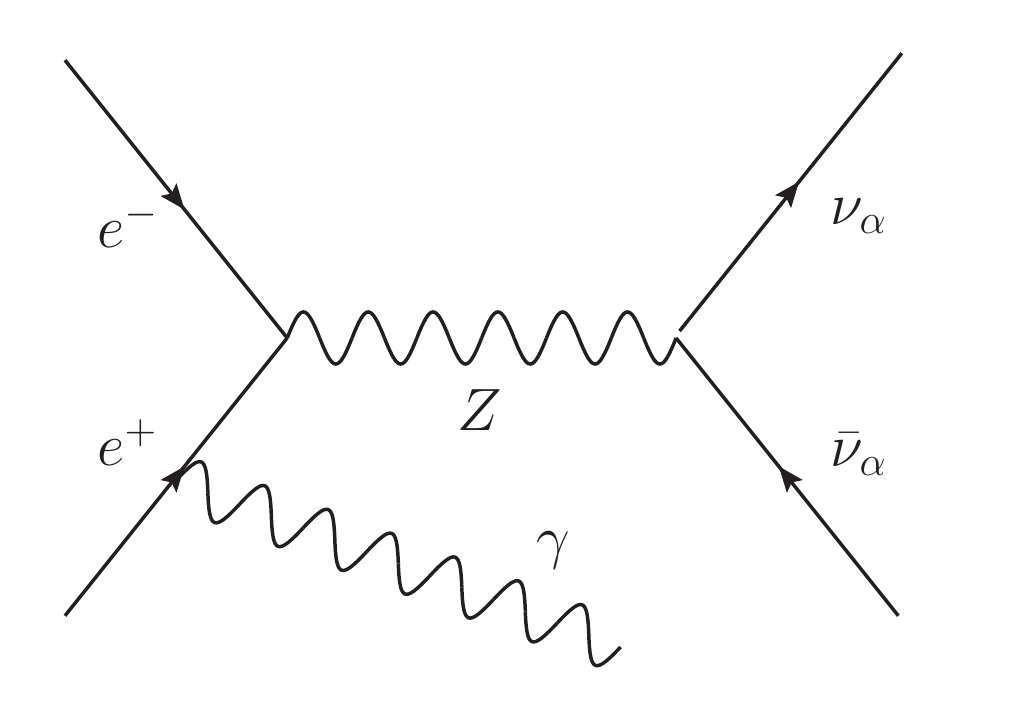}
	\end{minipage}
	\caption{Contributions to the $e^-e^+ \to \nu\bar{\nu}\gamma$ process at tree level, from the $W$ (a) and $Z$ (b) bosons.}
	\label{fig:feynDiag_nng}
\end{figure}

To compute the GNI contributions for this observable, we must consider
that its cross-section is written as the product~\cite{Nicrosini:1988hw, Barranco:2007ej,Berezhiani:2001rs}
\begin{equation}\label{eq:xsec_nunug}
	\frac{d^2\sigma}{dx\,dy}= H(x,y;s)\, \sigma_0(s(1 -x)),
\end{equation}
where $H(x,y;s)$ is the radiator function 
\begin{equation}
	H(x,y;s) = \frac{2\alpha \left[(1-\frac{1}{2}x)^2 + \frac{1}{4}x^2y^2\right]}{\pi x(1-y^2)},
\end{equation}
with
\begin{equation}
	x = 2 E_\gamma/\sqrt{s}, \quad \quad y = \cos\theta_{\gamma}, 
\end{equation}
where $\sqrt{s}$ is the center-of-mass energy and $E_\gamma (\theta_\gamma)$ is the energy (angle) of the produced photon. The term $\sigma_0$ is the ``reduced'' cross-section for  $e^+e^-
\to \nu\bar{\nu}$,
\begin{align}\label{eq:reducedXS}
	\sigma_0(s) &= \sigma_{W}(s) + \sigma_{Z}(s) + \sigma_{W-Z}(s), \nonumber \\
	\sigma_0(s) &= \frac{{G_F}^2 s}{12 \pi} \left[2 + \frac{N_{\nu} (g_V^2 + g_A^2)}{(1 - s / {M_Z}^2)^2 + \Gamma_Z^2 / M_Z^2} + \frac{2 (g_V + g_A) (1 - s / {M_Z}^2)}{(1 - s / M_Z^2)^2 + \Gamma_Z^2 / M_Z^2}\right] ,
\end{align}
where $\Gamma_Z$ and $M_Z$ are the $Z$ boson decay width and mass, respectively, and $N_\nu$ is the number of neutrino species. 
The first term in the previous equation comes from the $W$
contribution to the cross-section while the second one from the $Z$
contribution, as is illustrated in the Feynman diagrams in Fig.~(\ref{fig:feynDiag_nng}). The third contribution appears from the interference between both
gauge bosons.

Considering only neutral current GNI, we notice that the corrections
to this interaction will come from diagrams that in the SM are
related to the $Z$ bosons (Fig.~\ref{fig:feynDiag_nng}). From
Eq.~\eqref{equation_1} we can obtain the GNI cross-section
contribution for the process 
\begin{equation}\label{equation_2}
	\sigma_{0}=\frac{G^{2}_{F}s}{48\pi}\sum_{\alpha,\beta}\left(8|\epsilon^{e,L}_{\alpha\beta}|^{2}+8|\epsilon^{e,R}_{\alpha\beta}|^{2}+3|\epsilon^{e,S}_{\alpha\beta}|^{2}+3|\epsilon^{e,P}_{\alpha\beta}|^{2}+32|\epsilon^{e,T}_{\alpha\beta}|^{2}\right)\,.
\end{equation}
In Eq.~\eqref{equation_2}, we can see that tensor interactions contribute one order of magnitude more to the cross-section than the equally-contributing scalar and pseudoscalar interactions.

With this information we can proceed to compute the total cross-section 
\begin{equation}
	\sigma(s)=\int_{x_\mathrm{min}}^1 dx \int_{-\cos\theta_{\mathrm{min}}}^{\cos\theta_{\mathrm{min}}} dy H(x,y;s)\, \sigma_0(s(1 -x)).
	\label{eq:nng-totalXS}
\end{equation}

In the case of energies above the $Z$ resonance, it is important to
consider finite distance effects on the $W$ propagator. They require
the substitution~\cite{Hirsch:2002uv,
	Barranco:2007ej,Berezhiani:2001rs}:
\begin{align}\label{eq:reducedCrossSec2}
	\sigma_W(s) \; & \to \; \; \sigma_{W}(s) F_{W}(s/M_W^2), \nonumber \\
	\sigma_{W-Z}(s)\;&\to\; \; \sigma_{W-Z}(s) F_{W-Z}(s/M_{W}^{2}),
\end{align}
with 
\begin{align}
	F_{W}(z) &= \frac{3}{z^3} \left[-2(z+1)\log(z+1)+z(z+2)\right], \nonumber \\
	F_{W-Z}(z) &= \frac{3}{z^3} \left[(z+1)^2\log(z+1)-z(3z/2+1)\right].
\end{align}

From Eq.~\eqref{equation_2}, we can notice that the GNI observable that can be studied from this process is the following
\begin{equation}
	|\eps_{all}^{e,Y}|^2 \equiv \sum_{\alpha,\beta}|\eps_{\alpha\beta}^{e,Y}|^2,
	\label{eq:observable_nunugamma}
\end{equation}
where $Y=S,P,T$. Given that a neutrino-antineutrino pair is produced, this signal provides access to all the electron flavor-diagonal and flavor-changing GNI parameters.

\subsection{Differential cross-section for the process $\nu_{\alpha}+e^{-}\rightarrow \nu_{\beta}+e^{-}$}
The scattering of muon neutrinos off electrons is a purely leptonic
process that has been measured~\cite{Hasert:1973ff} since the
formulation of the SM. The small cross-section in comparison with the
scattering off nuclei makes it challenging to have high-statistics
samples. However, the clean leptonic signal yields independent
information complementary to that of the quark sector.

For electron neutrinos, it is difficult to have clean neutrino fluxes
at accelerator or spallation neutron sources. Therefore, the
measurement of this cross-section has been done mainly with reactor
antineutrino sources~\cite{Reines:1976pv,Daraktchieva:2005kn,Vidyakin:1992nf,Derbin:1993wy} that have even lower statistics. 

From Eq.~\eqref{equation_1}, we obtain the following result for
the differential cross-section for the process
$\nu_{\alpha}+e^{-}\rightarrow \nu_{\beta}+e^{-}$,~\cite{Bischer:2018zcz}
\begin{equation}\label{equation_5}
	\frac{d\sigma}{dE_r}=\frac{G^{2}_{F}m_{e}}{\pi}\left[A+2B\left(1-\frac{E_r}{E_\nu}\right)+C\left(1-\frac{E_r}{E_\nu}\right)^{2}+D\frac{m_{e}E_r}{E_\nu^{2}}\right]\,,
\end{equation}
where $E_r$ is the kinetic energy of the recoiling electron and
\begin{eqnarray}\label{equation_6}
	A&=&2|\epsilon^{e,L}_{\alpha\beta}|^{2}+\frac{1}{4}(|\epsilon^{e,S}_{\alpha\beta}|^{2}+|\epsilon^{e,P}_{\alpha\beta}|^{2})+8|\epsilon^{e,T}_{\alpha\beta}|^{2}-2\Re((\epsilon^{e,S}+\epsilon^{e,P})_{\alpha\beta}\epsilon^{e,T*}_{\alpha\beta})\,,\nonumber\\
	B&=&-\frac{1}{4}(|\epsilon^{e,S}_{\alpha\beta}|^{2}+|\epsilon^{e,P}_{\alpha\beta}|^{2})+8|\epsilon^{e,T}_{\alpha\beta}|^{2}\,,\nonumber\\
	C&=&2|\epsilon^{e,R}_{\alpha\beta}|^{2}+\frac{1}{4}(|\epsilon^{e,S}_{\alpha\beta}|^{2}+|\epsilon^{e,P}_{\alpha\beta}|^{2})+8|\epsilon^{e,T}_{\alpha\beta}|^{2}+2\Re((\epsilon^{e,S}+\epsilon^{e,P})_{\alpha\beta}\epsilon^{e,T*}_{\alpha\beta})\,,\nonumber\\
	D&=&-2\Re(\epsilon^{e,L}_{\alpha\beta}\epsilon^{e,R*}_{\alpha\beta})+\frac{1}{2}|\epsilon^{e,S}_{\alpha\beta}|^{2}-8|\epsilon^{e,T}_{\alpha\beta}|^{2}\,.
\end{eqnarray}
For the case of antineutrinos, the interchange $A\leftrightarrow  C$ has to be made in Eq.~\eqref{equation_5}. 
From the definitions in Eq.~\eqref{equation_6},
we notice that scalar and pseudoscalar interactions contribute equally
to the cross-section, except for the last term, D, where there is only
scalar contribution.  Given that this contribution is inversely
proportional to $E_\nu^2$ (last term of Eq.~\eqref{equation_5}), we
could distinguish between scalar and pseudoscalar parameters in
low-energy neutrino experiments.

Since the final (anti)neutrino flavor is unknown in an actual experiment, one has to sum over all the possible flavor outcomes. For this reason, the accessible observable for this process is
\begin{equation}
	|\eps_\alpha^{e,Y}|^2 \equiv \sum_\beta
	|\eps_{\alpha\beta}^{e,Y}|^2.  \label{eq:observable_nueScattering}
\end{equation}

\subsection{Neutrino-quark scattering}
Besides neutrino-electron scattering, neutrino-quark scattering gives
higher statistics
measurements~\cite{Dorenbosch:1986tb,Allaby:1987vr,Blondel:1989ev,Zeller:2001hh},
although one must take the uncertainties due to sea-quark corrections
with care.

We will analyze the cross-section for the process $\nu
q\rightarrow \nu q$.  The following expressions give the
neutrino-nucleon scattering cross-sections for the charged and neutral
weak currents in the SM~\cite{Zyla:2020zbs}
\begin{equation}
	\sigma^{CC}_{\nu N,SM}=\frac{G^{2}_{F}s}{2\pi}\left[f_{q}+\frac{1}{3}f_{\bar{q}}\right]\,,
	\label{eq:nuQuarkCrossSec_1} 
\end{equation}
\begin{equation}
	\begin{split}
		\sigma^{\nu N}_{NC,SM}=&\frac{G^{2}_{F}s}{2\pi}\Big[\left((g^{u,L})^{2}+\frac{1}{3}(g^{u,R})^{2}\right)f_{q}+\left((g^{d,L})^{2}+\frac{1}{3}(g^{d,R})^{2}\right)f_{q}\\
		&+\left((g^{u,R})^{2}+\frac{1}{3}(g^{u,L})^{2}\right)f_{\bar{q}}+\left((g^{d,R})^{2}+\frac{1}{3}(g^{d,L})^{2}\right)f_{\bar{q}}\Big]\,,
	\end{split}
	\label{eq:nuQuarkCrossSec_2}
\end{equation}
and for the antineutrino-nucleon scattering, we have the analog formulas
\begin{equation}
	\sigma^{CC}_{\bar{\nu} N,SM}=\frac{G^{2}_{F}s}{2\pi}\left[\frac{1}{3}f_{q}+f_{\bar{q}}\right]\,,
	\label{eq:nuQuarkCrossSec_1p} 
\end{equation}
\begin{equation}
	\begin{split}
		\sigma^{\bar{\nu} N}_{NC,SM}=&\frac{G^{2}_{F}s}{2\pi}\Big[\left((g^{u,R})^{2}+\frac{1}{3}(g^{u,L})^{2}\right)f_{q}+\left((g^{d,R})^{2}+\frac{1}{3}(g^{d,L})^{2}\right)f_{q}\\
		&+\left((g^{u,L})^{2}+\frac{1}{3}(g^{u,R})^{2}\right)f_{\bar{q}}+\left((g^{d,L})^{2}+\frac{1}{3}(g^{d,R})^{2}\right)f_{\bar{q}}\Big]\,,
	\end{split}
	\label{eq:nuQuarkCrossSec_2p}
\end{equation}
where the SM effective couplings $g^{f,L}$ and $g^{f,R}$ are:
\begin{equation}
	g^{f,L}=T^{f}_{3}-Q_{f}\sin^{2}\theta_{W}\,,\,\,\,\,\,\,g^{f,R}=-Q_{f}\sin^{2}\theta_{W}.
\end{equation}
Following Ref. \cite{Erler:2013xha}, we have,
\begin{equation}
	g^{u,L}=0.3457\,,\,\,\,\,g^{u,R}=-0.1553\,,\,\,\,\,g^{d,L}=-0.4288\,,\,\,\,\,g^{d,R}=0.0777\,.
\end{equation}
Furthermore, assuming an isoscalar target, $f_{q}$ and $f_{\bar{q}}$ are nuclear PDFs that regulate the fraction of proton momentum carried by the up and down quarks and anti-quarks, respectively. Neglecting the contributions of heavy quarks, we have $f_{q}\equiv f_{d}+f_{u}=\int_{0}^{1}x[u(x)+d(x)]dx$, and $f_{\bar{q}}=f_{\bar{d}}+f_{\bar{u}}=\int_{0}^{1}x[\bar{u}(x)+\bar{d}(x)]dx$.\\
The modifications introduced by the presence of scalar, pseudoscalar, and tensor interactions are given by \cite{Han:2020pff} 
\begin{equation}
	\sigma^{NC}_{\nu N,S(P)}=\sigma^{NC}_{\bar{\nu} N,S(P)}=\frac{G^{2}_{F}s}{24\pi}\left[\left(\epsilon^{u,S(P)}_{\alpha}\right)^{2}\left(\frac{f_{q}+f_{\bar{q}}}{2}\right)+\left(\epsilon^{d,S(P)}_{\alpha}\right)^{2}\left(\frac{f_{q}+f_{\bar{q}}}{2}\right)\right]\,,
	\label{eq:nuQuarkCrossSec_3}
\end{equation} 
\begin{equation}
	\sigma^{NC}_{\nu N,T}=\sigma^{NC}_{\bar{\nu} N,T}=\frac{28G^{2}_{F}s}{3\pi}\left[\left(\epsilon^{u,T}_{\alpha}\right)^{2}\left(\frac{f_{q}+f_{\bar{q}}}{2}\right)+\left(\epsilon^{d,T}_{\alpha}\right)^{2}\left(\frac{f_{q}+f_{\bar{q}}}{2}\right)\right]\, ,
	\label{eq:nuQuarkCrossSec_4}
\end{equation} 
where we have defined the effective couplings with quarks
\begin{equation}
	|\eps_\alpha^{q,Y}|^2 \equiv \sum_\beta |\eps_{\alpha\beta}^{q,X}|^2.
	\label{eq:observable_nuquarkScattering}
\end{equation}

\section{Experimental observables} 
\label{sec:experimental}

There is a large diversity of neutrino experiments that can provide
constraints to GNI couplings. In the present work we focus on the
``neutrino counting'' experiments ALEPH~\cite{Barate:1997ue,Barate:1998ci,Heister:2002ut}, DELPHI~\cite{Abreu:2000vk}, L3~\cite{Acciarri:1997dq,Acciarri:1998hb,Acciarri:1999kp}, and OPAL~\cite{Ackerstaff:1997ze,Abbiendi:1998yu,Abbiendi:2000hh}, the
accelerator experiments CHARM~\cite{Dorenbosch:1986tb,Allaby:1987vr}, CHARM-II~\cite{Vilain:1993kd,Vilain:1994qy}, CDHS~\cite{Blondel:1989ev}, and NuTeV~\cite{Zeller:2001hh}, and the
reactor experiment TEXONO~\cite{Deniz:2009mu}. These experiments can probe different GNI
couplings depending on the involved neutrino flavor and which fermion
interacts with them. In the following, we will separate the
experiments mentioned above according to the underlying process while
describing the implemented analysis and the fitted observable.

\subsection{Electron-positron collision}
A neutrino-antineutrino pair, along a single photon, can be produced 
in electron-positron collisions, as can be seen in the diagrams of Fig.~\ref{fig:feynDiag_nng}. Therefore, single-photon measurements can be used to derive limits to 
additional neutrino interactions. The LEP experiments ALEPH, DELPHI, 
L3, and OPAL had reported single-photon measurements at different 
energies above the $W^-W^+$ production threshold. In Table~\ref{Tab:experimentsAttributes} 
we present the required data to perform the analysis. In order of appearance,
each column consists of the center-of-mass energy, measured cross-section, 
expected cross-section reported by each collaboration, 
number of events after background subtraction, and the kinematic
cuts on energy and angle of the produced photon. For the last two columns,
$x_T = x\sin\theta_\gamma$, $x = E_\gamma/E_\mathrm{beam}$, and $y = \cos\theta_\gamma$.

\begin{table}[h]
	\centering
	\begin{tabular}{ccc@{\hskip 0.2in}c@{\hskip 0.3in}c@{\hskip 0.3in}c@{\hskip 0.3in}c@{\hskip 0.3in}c} \hline
		\toprule				
		& & $\sqrt{s}$ (GeV) & $\sigma^\mathrm{meas}$ (pb) & $\sigma^\mathrm{MC}$ (pb)
		& $N_\mathrm{obs}$ & $E_\gamma$ (GeV) & $|y|$ \\ \hline \hline
		\cline{1-8}      
		\multirow{10}{*}{ALEPH} & \multirow{2}{*}{\cite{Barate:1997ue}} & 161  & $5.3\pm0.83$ & $5.81\pm0.03$ & 41 & $x_T\geq 0.075$ & $\leq 0.95$ \\
		& & 172  & $4.7\pm0.83$ & $4.85\pm0.04$ & 36 & $x_T\geq 0.075$ & $\leq 0.95$ \\
		\cline{2-8}
		& \cite{Barate:1998ci} & 183  & $4.32\pm0.34$ & $4.15\pm0.03$ & 195 & $x_T\geq 0.075$ & $\leq 0.95$ \\
		\cline{2-8}
		& \multirow{7}{*}{\cite{Heister:2002ut}} & 189  & $3.43\pm 0.17$ & $3.48\pm 0.05$ & 484  & 
		\multirow{8}{*}{$x_T \geq 0.075$} & \multirow{8}{*}{$\leq 0.95$}\\
		& & 192  & $3.47\pm 0.40$ & $3.23\pm 0.05$ & 81  & & \\
		& & 196  & $3.03\pm 0.23$ & $3.26\pm 0.05$ & 197 & & \\
		& & 200  & $3.23\pm 0.22$ & $3.12\pm 0.05$ & 231 & & \\
		& & 202  & $2.99\pm 0.29$ & $3.07\pm 0.05$ & 110 & & \\
		& & 205  & $2.84\pm 0.22$ & $2.93\pm 0.05$ & 182 & & \\
		& & 207  & $2.67\pm 0.17$ & $2.80\pm 0.05$ & 292 & & \\
		\cline{1-8}
		\multirow{3}{*}{DELPHI} & \multirow{3}{*}{\cite{Abreu:2000vk}}	& 189  & $1.80\pm 0.20$  & $1.97$  & 146 & $x\geq 0.06$ & $\leq 0.7$ \\
		& & 183  & $2.33\pm 0.36$  & $2.08$  & 65 & $x\geq 0.02$ & $\leq 0.85$ \\
		& & 189  & $1.89\pm 0.22$  & $1.94$  & 155 & $x\leq 0.9$	& $\leq 0.98$ \\
		\cline{1-8}
		\multirow{6}{*}{L3}	& \multirow{3}{*}{\cite{Acciarri:1997dq}}	& 161  & $6.75\pm 0.93$  & $6.26\pm 0.12$  & 57 & $\geq 10$ &  $\leq 0.73$  \\
		& & & & & & and & and \\
		& & 172  & $6.12\pm 0.90$  & $5.61\pm 0.10$  & 49 & $E_T\geq 6$ & 0.80-0.97 \\
		\cline{2-8}		
		& \cite{Acciarri:1998hb} & 183  & $5.36\pm 0.40$  & $5.62\pm 0.10$  & 195 &$\geq 5$ &  $\leq 0.73$  \\
		& & & & & & and & and \\
		& \cite{Acciarri:1999kp} & 189  & $5.25\pm 0.23$  & $5.29\pm 0.06$  & 572 &  $E_T\geq 5$ & 0.80-0.97 \\
		\cline{1-8}
		\multirow{10}{*}{OPAL} & \multirow{3}{*}{\cite{Ackerstaff:1997ze}}	& 130  & $10.0\pm 2.34$  & $13.48\pm0.22$ & 19  & $x_T > 0.05$ &  $\leq 0.82$  \\
		& & & & & & and & and \\
		& & 136  & $16.3\pm 2.89$  & $11.30\pm 0.20$  & 34  & $x_T > 0.1$ &  $\leq 0.966$ \\
		\cline{2-8}
		& \multirow{2}{*}{\cite{Abbiendi:1998yu}} & 130  & $11.6\pm 2.53$  & $14.26\pm 0.06$  & 21  & \multirow{2}{*}{$x_T > 0.05$} &  \multirow{2}{*}{$\leq 0.966$}\\
		& & 136  & $14.9\pm 2.45$  & $11.95\pm 0.07$  & 39  &  &  \\
		\cline{2-8}
		& \multirow{3}{*}{\cite{Ackerstaff:1997ze}} & 161  & $5.30\pm 0.83$  & $6.49\pm 0.08$  & 40  & $x_T > 0.05$ &  $\leq 0.82$  \\
		& & & & & & and & and \\
		& & 172  & $5.50\pm 0.83$  & $5.53\pm 0.08$  & 45 & $x_T > 0.1$ & $\leq 0.966$\\
		\cline{2-8}
		& \cite{Abbiendi:1998yu} & 183  & $4.71\pm 0.38$  & $4.98\pm 0.02$  & 191 & $x_T > 0.05$ & $\leq 0.966$\\
		\cline{2-8}
		& \cite{Abbiendi:2000hh} & 189  & $4.35\pm 0.19$  & $4.66\pm 0.03$  & 643  & $x_T > 0.05$ & $\leq 0.966$ \\			          
		\hline
	\end{tabular}
	\caption{Data summary from the ALEPH, DELPHI, L3, and OPAL experiments.}
	\label{Tab:experimentsAttributes}
\end{table}

The constraints on different GNI couplings can be obtained by performing a $\chi^2$ analysis
with the function
\begin{equation}
	\chi^2_{e^-e^+} = \sum_i \left(\frac{\sigma_i^\mathrm{th} - \sigma_i^\mathrm{meas}}{\sigma_i}\right)^2.
\end{equation}
Here, $\sigma_i^\mathrm{th}$ is the theoretical cross-section
including new interactions, which can be computed by integrating
Eq.~\eqref{eq:nng-totalXS} in the limits shown in the last two columns
of Table~\ref{Tab:experimentsAttributes}. The measured cross-section
$\sigma_i^\mathrm{meas}$ and its uncertainty $\sigma_i$ can be read
from the second column of Table~\ref{Tab:experimentsAttributes}. The
subscript $i$ stands for the different measurements of each
experiment. Given the details of the experimental cuts, we include a
normalization factor with a $10\%$ systematic uncertainty in order to
improve the robustness of the analysis~\cite{Escrihuela:2019mot}.
  
\subsection{Neutrino-electron scattering}

The CHARM-II experiment has measured the 
neutrino-electron scattering cross-section 
from a purely neutral-current
interaction~\cite{Vilain:1993kd,Vilain:1994qy}. A total of $2677 \pm
82 $ $\nu_\mu e^-$ and $2752 \pm 88$ $\bar{\nu}_\mu e^-$ events were
detected using $\nu_\mu$ and $\bar{\nu}_\mu$ beams produced at CERN,
with mean energies of 23.7 and 19.1 GeV, respectively.

To perform a fit over the GNI parameters, we use only the spectral information on the cross section reported by
	the experiment. We define the following function
\begin{equation}
	\chi^2_\mathrm{CHARM-II} = \sum_{i=1}^8 \left(\frac{(\tfrac{d\sigma}{dT})_i^\mathrm{th} - (\tfrac{d\sigma}{dT})_i^\mathrm{meas}}{\sigma_i}\right)^2,
\end{equation} 
where $(\tfrac{d\sigma}{dT})_i^\mathrm{th}$ is the theoretical cross-section 
(including GNI) from Eq.~\eqref{equation_5}, $(\tfrac{d\sigma}{dT})_i^\mathrm{meas}$ is the experimental
measurement, and $\sigma_i$ its uncertainty~\cite{Vilain:1993kd}. The subscript $i$ runs over all the
neutrino (one to four) and antineutrino (five to eight) measurements.

The TEXONO collaboration has measured the $\bar{\nu}_e e^-$ scattering
cross section using a CsI(Tl) scintillating crystal detector placed
near the reactor at the Kuo-Sheng Nuclear Power
Plant~\cite{Deniz:2009mu}. Events in the $3-8$~MeV recoil energy range
were reported, resulting in 414~$\pm$~80(stat)~$\pm$~61(syst)
events in ten electron recoil energy bins, after background subtraction.

In order to extract limits for GNI, we use again only the event rate spectral information. We adopt the following least-squares function
\begin{equation}
	\chi^2_\mathrm{TEXONO} = \sum_{i=1}^{10} \left(\frac{N_i^\mathrm{th} - N_i^\mathrm{meas}}{\sigma_i}\right)^2,
\end{equation}
where $N_i^\mathrm{th}$ is the theoretical number of events, while 
 $N_i^\mathrm{meas}$ refers to the measured number 
of events in the $i$-th recoil energy bin, and $\sigma_i$ is the measured uncertainty~\cite{Deniz:2009mu}. The theoretical number of events is computed according to
\begin{equation}
	N_i^\mathrm{th} = A\int_{E_r^i}^{E_r^{i+1}}\int_{E_\nu^\mathrm{min}}^{E_\nu^\mathrm{max}} \phi(E_\nu) \frac{d\sigma}{dE_r}(E_r,E_\nu)\,dE_\nu \, dE_r,
	\label{eq:eventsTexono}
\end{equation}
with $d\sigma/dE_r$ as the differential cross-section in Eq.~\eqref{equation_5}, $\phi(E_\nu)$ as
the reactor $\bar{\nu}_e$ energy spectrum, and A stands for the product of the number of targets, total neutrino flux, and running time of the experiment~\cite{Deniz:2009mu}. We consider the theoretical 
prediction of the Huber+Mueller model for the reactor spectrum, since it spans energies
in the $2-8$ MeV range~\cite{Huber:2011wv,Mueller:2011nm}. In the first integration of 
Eq.~\eqref{eq:eventsTexono}, the upper limit is the maximum available energy from the
reactor, which for our case is $E_\nu^\mathrm{max} = 8$~MeV, while the lower limit 
$E_\nu^\mathrm{min}$ can be extracted from the kinematic relation
\begin{equation}
	T^\mathrm{max} = \frac{2E_\nu^2}{m_e + 2E_\nu}.
\end{equation}

Additionally, the Borexino experiment has measured solar neutrinos
  from the $pp$, $^7$Be, $pep$, $^8$B
  fluxes~\cite{Bellini:2011rx,Collaboration:2011nga,Bellini:2014uqa,
    Agostini:2017ixy, Agostini:2017cav}, and CNO
  cycle~\cite{Agostini:2020mfq}, through their elastic scattering off
  electrons. These measurements can also be used to probe GNI
  parameters, and might help to improve the constrains on pseudoscalar
  interactions, given the low energy of solar neutrinos (see
  discussion in previous section).  However, a detailed study of
  this data set should include several systematic uncertainties;
  therefore, we will take a more conservative approach and we will not
  consider the Borexino measurements in our analysis.  

\subsection{Deep inelastic scattering}
Neutrino deep inelastic scattering with nucleons can constrain generalized neutrino interactions with quarks. 
For this purpose we consider measurements from the
CHARM~\cite{Dorenbosch:1986tb,Allaby:1987vr}, CDHS~\cite{Blondel:1989ev}, 
and NuTeV~\cite{Zeller:2001hh} experiments.

By using an electron-neutrino beam with equal $\nu_e$ and $\bar{\nu}_e$ fluxes,
the CHARM collaboration measured the ratio of total cross-section for semileptonic 
scattering defined as~\cite{Dorenbosch:1986tb}
\begin{equation}
	R^e \equiv \frac{\sigma(\nu_e N \to \nu X) + \sigma(\bar{\nu}_e N \to \nu X)}{\sigma(\nu_e N \to e^- X) + \sigma(\bar{\nu}_e N \to e^+ X)}. 
\end{equation}

The reported value is $R^e = 0.406 \pm 0.140$~\cite{Dorenbosch:1986tb}, while the SM prediction
following Eqs.~\eqref{eq:nuQuarkCrossSec_1} and \eqref{eq:nuQuarkCrossSec_2} is~\cite{Zyla:2020zbs} 
\begin{equation}
	R^e_\mathrm{SM} = g_{L}^2 + g_{R}^2 = 0.335.
\end{equation}
When including new interactions, this ratio takes the form~\cite{Han:2020pff} 
\begin{equation}
	R^e = g_{L}^2 + g_{R}^2 + \frac{1}{12}\sum_{q=u,d}\left(|\eps^{q,S}_e|^2 + |\eps^{q,P}_e|^2 + 224 |\eps^{q,T}_e|^2 \right),
\end{equation}
where $|\eps^{q,Y}_e|^2$ are the effective couplings defined in Eq.~\eqref{eq:observable_nuquarkScattering}.
A simple $\chi^2$ function for the fit can be expressed as
\begin{equation}
	\chi^2_{\mathrm{CHARM}-e} = \left(\frac{R^{e,\mathrm{th}} - R^e}{\sigma^e}\right)^2.
\end{equation}

On the other hand, the accelerator experiments CHARM, CDHS, and NuTeV measured the ratios
of neutral to charged current semileptonic cross-sections $R^\nu$ and $R^{\bar{\nu}}$
using a muon-neutrino beam. The reported values are shown in Table~\ref{Tab:DIS_measurements}.

\begin{table}[ht]
	\centering
	\begin{tabular}{c@{\hskip 0.2in}c@{\hskip 0.2in}c@{\hskip 0.2in}c} \hline
		\toprule
		& $R^\nu$ & $R^{\bar{\nu}}$ & $r$  \\   \hline  \hline
		CHARM~\cite{Allaby:1987vr} & $0.3093 \pm 0.0031$ & $0.390 \pm 0.014$ & $0.456 \pm 0.011$ \\
		CDHS~\cite{Blondel:1989ev} & $0.3072 \pm 0.0033$ & $0.382 \pm 0.016$ & $0.393 \pm 0.014$ \\
		NuTeV~\cite{Bentz:2009yy} & $0.3933 \pm 0.0015$ & $0.4034 \pm 0.0028$ & -- \\  \hline
	\end{tabular} 
	\caption{\footnotesize Values of the neutral to charged current cross 
		section ratios $R^\nu$ and 
		$R^{\bar{\nu}}$ reported by the different experiments. The CHARM and CDHS
		collaborations also report a value for the ratio $r$ defined in Eq.~\eqref{eq:parameter_r}.
		The numbers shown for the NuTeV experiment are extracted from Ref.~\cite{Bentz:2009yy}
		(see text for details). } 
	\label{Tab:DIS_measurements}    
\end{table}

These ratios are defined as~\cite{Jonker:1980vf}
\begin{equation}
	R^\nu \equiv \frac{\sigma(\nu_\mu N \to \nu X)}{\sigma(\nu N \to \mu^- X)}, \quad 	R^{\bar{\nu}} \equiv \frac{\sigma(\bar{\nu}_\mu N \to \bar{\nu} X)}{\sigma(\bar{\nu} N \to \mu^+ X)} .
\end{equation}
Again, with the help of Eqs.~\eqref{eq:nuQuarkCrossSec_1} and \eqref{eq:nuQuarkCrossSec_2}, we can
find the expressions of $R^\nu$ and $R^{\bar{\nu}}$ in the SM framework,
\begin{equation}
	R^\nu_\mathrm{SM} = g_L^2 + r g_R^2, \quad R^{\bar{\nu}}_\mathrm{SM} = g_L^2 + \frac{1}{r} g_R^2. 
\end{equation}
Using Eqs.~\eqref{eq:nuQuarkCrossSec_3} and
\eqref{eq:nuQuarkCrossSec_4}, we can compute the equivalent expressions for the GNI case~\cite{Han:2020pff}
\begin{eqnarray}
	R^\nu = g_L^2 + r g_R^2 + \frac{1}{32}(1+r)\sum_{q=u,d}\left(|\eps^{q,S}_\mu|^2 + |\eps^{q,P}_\mu|^2 + 224 |\eps^{q,T}_\mu|^2 \right), \nonumber \\
	R^{\bar{\nu}} = g_L^2 + \frac{1}{r} g_R^2 +  \frac{1}{32}(1+\frac{1}{r})\sum_{q=u,d}\left(|\eps^{q,S}_\mu|^2 + |\eps^{q,P}_\mu|^2 + 224 |\eps^{q,T}_\mu|^2 \right).
	\label{eq:ratiosGNI_CHARM_CDHS}
\end{eqnarray}
The parameter $r$ from above corresponds to the ratio of charged-current cross-sections~\cite{Jonker:1980vf}
\begin{equation}
	r = \frac{\sigma(\bar{\nu} N \to \mu^+ X)}{\sigma(\nu N \to \mu^- X)}
	\label{eq:parameter_r} = \frac{\tfrac{1}{3}f_q + f_{\bar{q}}}{f_q + \tfrac{1}{3}f_{\bar{q}}} .
\end{equation} 
It is worth noticing that in the scope of this work, the ratio $r$ is unaffected by the 
presence of GNI terms since we only considered neutral-current interactions.
Given that $r$ appears in both expressions from Eq.~\eqref{eq:ratiosGNI_CHARM_CDHS}, we have adopted a $\chi^2$ function that includes correlation
between the neutrino and antineutrino ratios, for both CHARM and CDHS analyses:
\begin{equation}
	\chi^2_{\mathrm A} = \sum_{j,k}(R^j_\mathrm{th} - R^j)(\sigma^2)^{-1}_{jk}(R^k_\mathrm{th} - R^k),
\end{equation}
where $\sigma^2$ is the covariance matrix constructed from the squared
errors, $R^j_\mathrm{th}$ is the ratio including new interactions,
$R^j$ is the measured one, and $j,k=\nu,\bar{\nu}$. The label
$\mathrm{A}$ refers to either $\mathrm{CHARM}$ or $\mathrm{CDHS}$.

For the case of NuTeV, there is not reported value for the ratio
$r$. Therefore, we will express $R^\nu$ and $R^{\bar{\nu}}$ in terms
of the fractions $f_q$ and $f_{\bar{q}}$ in both the SM~\cite{Zyla:2020zbs}
\begin{equation}
	R^\nu_\mathrm{SM} = g_L^2 + \frac{\tfrac{1}{3}f_q + f_{\bar{q}}}{f_q + \tfrac{1}{3}f_{\bar{q}}} g_R^2, \quad R^{\bar{\nu}}_\mathrm{SM} = g_L^2 + \frac{f_q + \tfrac{1}{3}f_{\bar{q}}}{\tfrac{1}{3}f_q + f_{\bar{q}}} g_R^2,
	\label{eq:ratiosSM_CHARM_CDHS_2}
\end{equation}
while in the GNI framework we have~\cite{Han:2020pff} 
\begin{eqnarray}
	R^\nu = R^\nu_\mathrm{SM} + \frac{1}{24}\left(\frac{f_q + f_{\bar{q}}}{f_q + \tfrac{1}{3}f_{\bar{q}}}\right)\sum_{q=u,d}[(\eps^{q,S}_\mu)^2 + (\eps^{q,P}_\mu)^2 + 224 (\eps^{q,T}_\mu)^2 ], \nonumber \\
	R^{\bar{\nu}} = R^{\bar{\nu}}_\mathrm{SM} +  \frac{1}{24}\left(\frac{f_q + f_{\bar{q}}}{\tfrac{1}{3}f_q + f_{\bar{q}}}\right)\sum_{q=u,d}[(\eps^{q,S}_\mu)^2 + (\eps^{q,P}_\mu)^2 + 224 (\eps^{q,T}_\mu)^2 ].
	\label{eq:ratiosGNI_CHARM_CDHS_2}
\end{eqnarray}
At $Q^2 = 20 \mathrm{GeV}^2$, we have $f_q = 0.42$ and $f_{\bar{q}} = 0.068$~\cite{Han:2020pff}. 

For this experiment we adopt the $\chi^2$ function
\begin{equation}
	\chi^2_\mathrm{NuTeV} = \sum_i \left(\frac{a^iR^i_\mathrm{th} - R^i}{\sigma_i}\right)^2,
	\label{eq:chi_nutev}
\end{equation}
with $a^i$ a normalization factor, $R^i_\mathrm{th}$ the ratio defined
in Eq.~\eqref{eq:ratiosGNI_CHARM_CDHS_2}, $R^i$ the experimental
measurement, $\sigma_i$ its uncertainty, and $i=\nu,\bar{\nu}$. In the
SM framework, using Eq.~\eqref{eq:ratiosSM_CHARM_CDHS_2}, we have
$R^\nu_\mathrm{SM} = 0.3175$ and $R^{\bar{\nu}}_\mathrm{SM} = 0.3675$,
which are in strong disagreement with the NuTeV measurements. However, after 
including corrections associated with charge symmetry violation,
momentum carried by s-quarks, and parton distribution functions the
SM predictions become $R^\nu_\mathrm{SM,\, corr} = 0.3950$ and
$R^{\bar{\nu}}_\mathrm{SM, \, corr} = 0.4066$, in better
agreement with the NuTeV reported values~\cite{Bentz:2009yy}.  Hence
we have chosen as normalization factors, $a^{i}$, 
\begin{equation}
	a^\nu = \frac{R^{\nu}_\mathrm{SM, \, corr}}{R^\nu_\mathrm{SM}}, \quad  a^{\bar{\nu}} = \frac{R^{\bar{\nu}}_\mathrm{SM, \, corr}}{R^{\bar{\nu}}_\mathrm{SM}}.
\end{equation}

By comparing the different measurements from Table III, NuTeV provided
a more accurate measurement, leading to more stringent bounds on new
physics. However, as stated before, the initial results from the NuTeV
collaboration differ by almost $3\sigma$ from the SM
prediction~\cite{Zeller:2001hh}. For this reason, in the following
section, we present the derived limits on the GNI couplings with and
without the NuTeV measurements.

\section{Results}
\label{sec:results}

In this section, we will summarize the results obtained from the
$\chi^2$ analysis of the experiments described in the previous
section, assuming the presence of new interactions. We derived limits
for the scalar, pseudoscalar, and tensor couplings following different
approaches. First, we considered data from each experiment at a time;
next, we performed a global fit considering several experiments to fit
the relevant parameters.

As mentioned in Section~\ref{sec:formalism}, each experiment is
sensitive to a different combination of GNI parameters. These
combinations, which we will now refer to as \emph{observables}, are
the effective GNI couplings we have defined in
Eqs.~\eqref{eq:observable_nunugamma},\eqref{eq:observable_nueScattering},
and~\eqref{eq:observable_nuquarkScattering}. Thus, each experiment
depends on three different observables (scalar, pseudoscalar, and
tensor).  We consider first only one of these observables different
from zero at a time.  In Table~\ref{tab:experimentsLimits}, we present
the limits, at $90\%$ CL, for the scalar and pseudoscalar parameters
obtained from the fit for each experiment. Note that these bounds are
the same ($\epsilon^{f,S}_{\alpha \beta}
= \epsilon^{f,P}_{\alpha \beta} $) for most of the experiments
considered in this work, given the symmetry of these terms in the
cross-sections. The only exception is the TEXONO experiment due to the
low energy of reactor antineutrinos (see discussion in
Section~\ref{sec:formalism}). The data from the $e^+e^-$ collision
also have particular relevance in our analysis, since it shows
sensitivity to the diagonal $\epsilon^{e,S,P}_{\tau \tau}$ parameter 
due to its inclusive character.

Additionally, we show the resulting limits for tensor interactions in
Table~\ref{tab:experimentsLimits2}, also at $90\%$ CL.  These limits
are more stringent than those for scalar and pseudoscalar interactions
because the tensor term in the different cross-sections is
approximately ten times higher. As expected, bounds coming from
muon-neutrino experiments are more restrictive than those from
electron-neutrino, given their higher statistics.

\begin{table}[ht]
	\centering
	\begin{tabular}{ccc@{\hskip 0.2in}c} \hline
		\toprule
		Experiment & Observable & Parameters & Limit  \\   \hline  \hline
		ALEPH~\cite{Barate:1997ue, Barate:1998ci, Heister:2002ut} & \multirow{4}{*}{$|\eps^{e,X}_{all}|$} & \multirow{4}{*}{$|\eps^{e,X}_{ee}|$, $|\eps^{e,X}_{\mu\mu}|$, $|\eps^{e,X}_{\tau\tau}|$, $|\eps^{e,X}_{e\mu}|$, $|\eps^{e,X}_{e\tau}|$, $|\eps^{e,X}_{\mu\tau}|$} & $<0.535$  \\ 
		DELPHI~\cite{Abreu:2000vk} &  &  & $<0.830$  \\  
		L3~\cite{Acciarri:1997dq, Acciarri:1998hb, Acciarri:1999kp} &  &  & $<0.745$  \\  
		OPAL~\cite{Ackerstaff:1997ze, Abbiendi:1998yu, Abbiendi:2000hh} &  &  & $<0.637$  \\  \hline
		CHARM-II~\cite{Vilain:1993kd} & $|\eps^{e,X}_{\mu}|$ & $|\eps^{e,X}_{e\mu}|$, $|\eps^{e,X}_{\mu\mu}|$, $|\eps^{e,X}_{\mu\tau}|$ &  $<0.401$ \\  \hline
		TEXONO~\cite{Deniz:2009mu} & $|\eps^{e,X}_{e}|$ & $|\eps^{e,X}_{ee}|$, $|\eps^{e,X}_{e\mu}|$, $|\eps^{e,X}_{e\tau}|$ &  $|\eps^{e,S}_{e}|<0.56$, \,\, $|\eps^{e,P}_{e}|<0.64$ \\  \hline
		CHARM~\cite{Dorenbosch:1986tb} (\stackon[.1pt]{$\nu$}{\brabar}$_e$ beam) & $|\eps^{q,X}_{e}|$ & $|\eps^{q,X}_{ee}|$, $|\eps^{q,X}_{e\mu}|$, $|\eps^{q,X}_{e\tau}|$ & $<1.9$  \\  \hline
		CHARM~\cite{Allaby:1987vr} (\stackon[.1pt]{$\nu$}{\brabar}$_\mu$ beam) & \multirow{3}{*}{$|\eps^{q,X}_{\mu}|$} & \multirow{3}{*}{$|\eps^{q,X}_{e\mu}|$, $|\eps^{q,X}_{\mu\mu}|$, $|\eps^{q,X}_{\mu\tau}|$} &  $<0.205$ \\ 
		CDHS~\cite{Blondel:1989ev} & & & $<0.198$\\ 
		NuTeV~\cite{Zeller:2001hh} & & & $<0.11 $\\  \hline
	\end{tabular} 
	\caption{\footnotesize Exclusion $90\%$ C.L. limits on the observable scalar  and pseudoscalar neutrino interaction parameter for different experiments, with $X=S,P$. Both scalar and pseudoscalar limits are the same for all experiments except for TEXONO (see Sec.~\ref{sec:formalism} for a detailed explanation). } \label{tab:experimentsLimits}   
\end{table}
%
\begin{table}[ht]
	\centering
	\begin{tabular}{ccc@{\hskip 0.2in}c} \hline
		\toprule
		Experiment & Observable & Parameters & Limit  \\   \hline  \hline
		ALEPH~\cite{Barate:1997ue, Barate:1998ci, Heister:2002ut} & \multirow{4}{*}{$|\eps^{e,T}_{all}|$} & \multirow{4}{*}{$|\eps^{e,T}_{ee}|$, $|\eps^{e,T}_{\mu\mu}|$, $|\eps^{e,T}_{\tau\tau}|$, $|\eps^{e,T}_{e\mu}|$, $|\eps^{e,T}_{e\tau}|$, $|\eps^{e,T}_{\mu\tau}|$} & $<0.163$  \\ 
		DELPHI~\cite{Abreu:2000vk} &  &  & $<0.254$  \\  
		L3~\cite{Acciarri:1997dq, Acciarri:1998hb, Acciarri:1999kp} &  &  & $<0.228$  \\  
		OPAL~\cite{Ackerstaff:1997ze, Abbiendi:1998yu, Abbiendi:2000hh} &  &  & $<0.194$  \\  \hline
		CHARM-II~\cite{Vilain:1993kd} & $|\eps^{e,T}_{\mu}|$ & $|\eps^{e,T}_{e\mu}|$, $|\eps^{e,T}_{\mu\mu}|$, $|\eps^{e,T}_{\mu\tau}|$ &  $<0.036$ \\  \hline
		TEXONO~\cite{Deniz:2009mu} & $|\eps^{e,T}_{e}|$ & $|\eps^{e,T}_{ee}|$, $|\eps^{e,T}_{e\mu}|$, $|\eps^{e,T}_{e\tau}|$ &  $<0.073$ \\  \hline
		CHARM~\cite{Dorenbosch:1986tb} (\stackon[.1pt]{$\nu$}{\brabar}$_e$ beam) & $|\eps^{q,T}_{e}|$ & $|\eps^{q,T}_{ee}|$, $|\eps^{q,T}_{e\mu}|$, $|\eps^{q,T}_{e\tau}|$ & $<0.127$  \\  \hline
		CHARM~\cite{Allaby:1987vr} (\stackon[.1pt]{$\nu$}{\brabar}$_\mu$ beam) & \multirow{3}{*}{$|\eps^{q,T}_{\mu}|$} & \multirow{3}{*}{$|\eps^{q,T}_{e\mu}|$, $|\eps^{q,T}_{\mu\mu}|$, $|\eps^{q,T}_{\mu\tau}|$} &  $<0.0137$ \\ 
		CDHS~\cite{Blondel:1989ev} & & & $<0.0130$\\ 
		NuTeV~\cite{Zeller:2001hh} & & & $<0.00754 $\\  \hline
	\end{tabular} 
	\caption{\footnotesize Exclusion $90\%$ C.L. limits on the observable tensor neutrino interaction parameter for different experiments.} \label{tab:experimentsLimits2}    
\end{table}

\begin{figure}
	\centering
	\includegraphics[width=\linewidth]{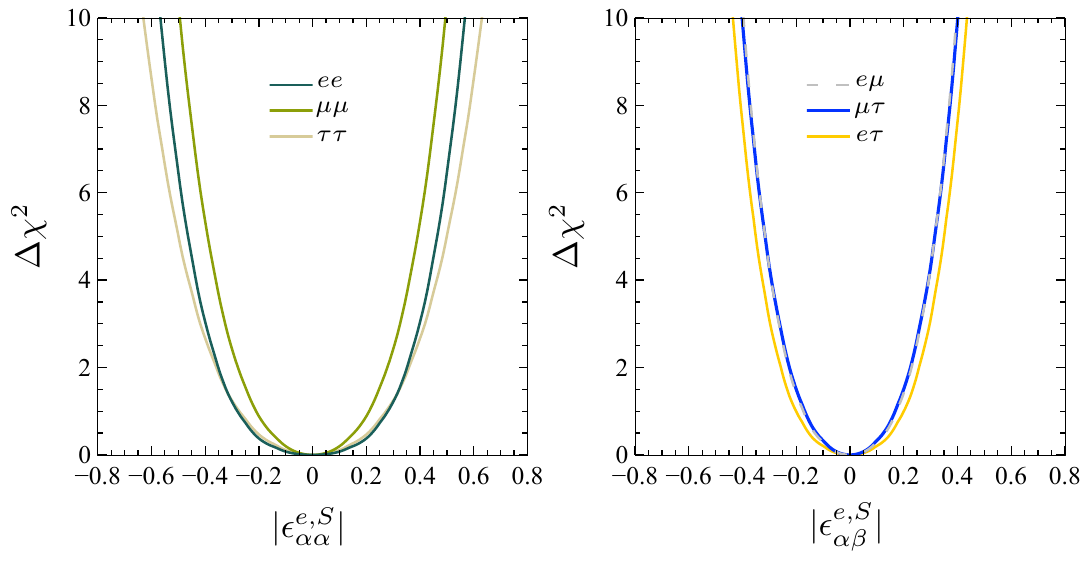}
	\includegraphics[width=\linewidth]{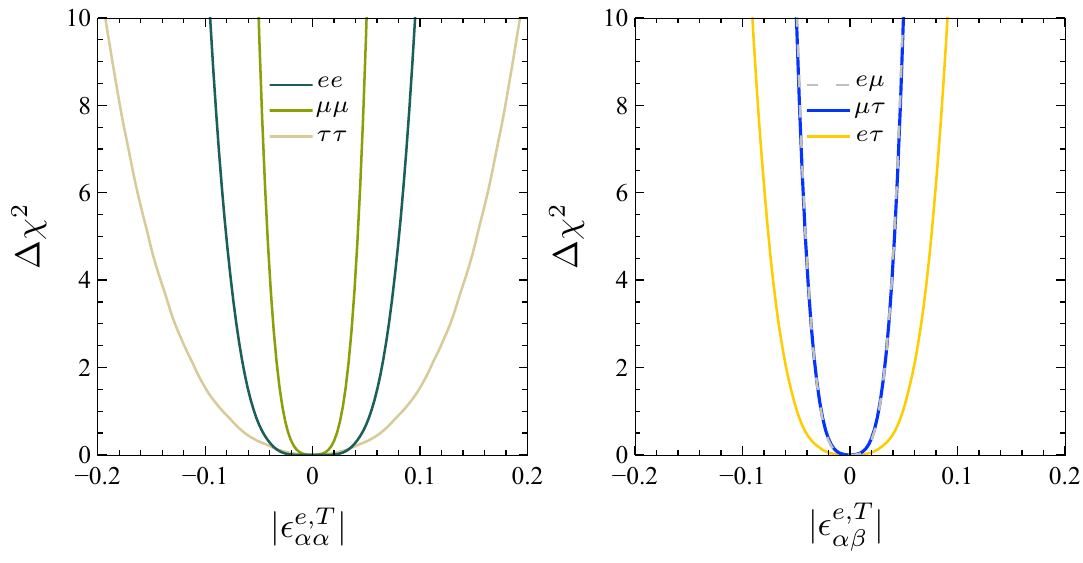}
	\caption{Global constraints on the neutrino-electron interaction parameters from different experiments. The upper (lower) panels correspond to the scalar (tensor) parameters. 
		In the left (right) panels we present limits for the flavor-diagonal (changing) parameters. }
	\label{fig:boundsExps_electron}
\end{figure}

The second part of the analysis shows a global study of the data from different experiments to obtain bounds for a specific GNI coupling. 
For example, for the parameter
$|\eps_{ee}^{e,Y}|$ we combine data from the $e^+e^-$ collision
experiments and TEXONO, and set every other GNI parameter equal to
zero. In this sense, we derive more restrictive limits than the presented in
Tables~\ref{tab:experimentsLimits}
and~\ref{tab:experimentsLimits2}. We show in
Fig.~\ref{fig:boundsExps_electron} the $\chi^2$ profile of the
different scalar and tensor GNI parameters from neutrino-electron
interactions. We have omitted the pseudoscalar parameters from this
plot due to their similarity with the scalar parameters. 

\begin{figure}
	\centering
	\includegraphics[width=\linewidth]{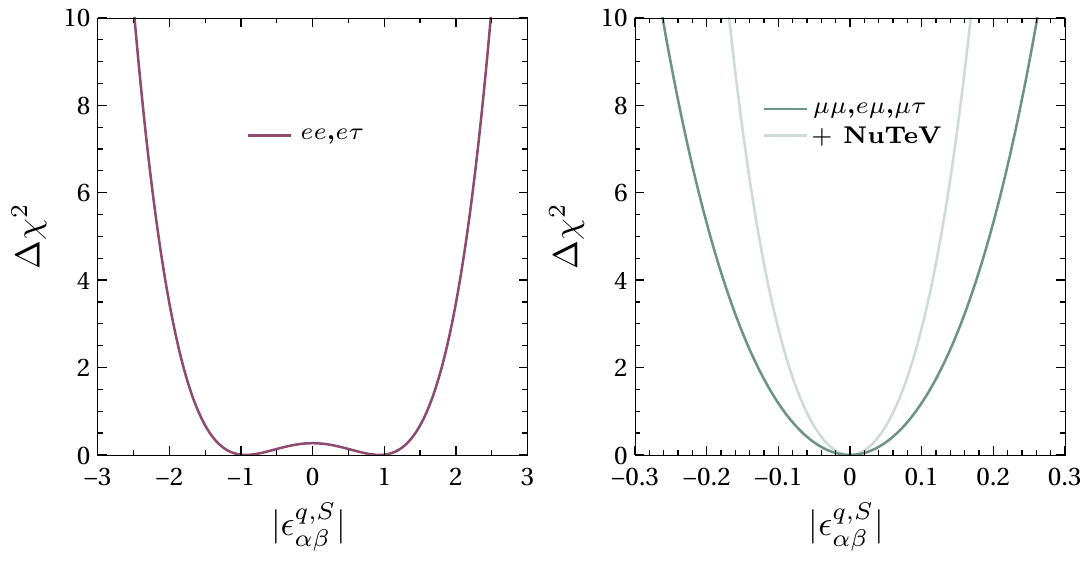}
	\includegraphics[width=\linewidth]{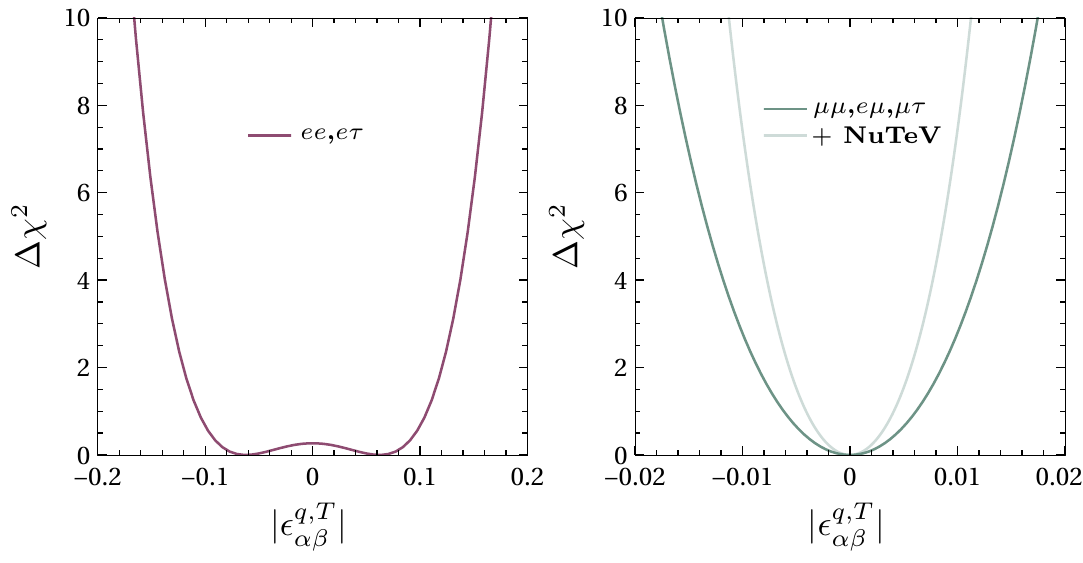}
	\caption{Global constraints on the neutrino-quark interaction parameters from different experiments. The upper (lower) panels correspond to the scalar (tensor) parameters. 
		In the right panels we present the resulting limits without the NuTeV measurement (solid green line) and including it (solid gray line).}
	\label{fig:boundsExps_quark}
\end{figure}

As for the neutrino-quark interactions, we introduce in
Fig.~\ref{fig:boundsExps_quark} the corresponding profiles, where we
have chosen to present the results with and without including data
from NuTeV. We can see here that the preferred value for
$|\eps^{q,Y}_{ee}|$ and $|\eps^{q,Y}_{e\tau}|$ is different from zero,
unlike every other GNI parameter studied in this work.

The resulting $90\%$~C.L. limits for all the GNI parameters, and the experiments used to derive them, are listed in 
Table~\ref{tab:combinedLimits}. As stated before, only the limits associated with the TEXONO experiment can differentiate between
scalar and pseudoscalar interactions. 

\begin{table}[ht]
	\centering
	\begin{tabular}{cc@{\hskip 0.3in}cc} \hline
		\toprule
		Experiments & Scalar &  Pseudoscalar  	&  Tensor  	\\   \hline  \hline
		$e^-e^+$ + TEXONO & $|\eps^{e,S}_{ee}| <0.38$ & $|\eps^{e,P}_{ee}|<0.40$	& $|\eps^{e,T}_{ee}|<0.07$	\\ 
		$e^-e^+$ + CHARM-II & \multicolumn{2}{c}{$|\eps^{e,X}_{\mu\mu}|<0.31$}  & $|\eps^{e,T}_{\mu\mu}|<0.03$	\\  
		$e^-e^+$ & \multicolumn{2}{c}{$|\eps^{e,X}_{\tau\tau}|<0.40$}	& $|\eps^{e,T}_{\tau\tau}|<0.12$	\\  
		$e^-e^+$ + TEXONO + CHARM-II & $|\eps^{e,S}_{e\mu}|<0.25$ & $|\eps^{e,P}_{e\mu}|<0.25$	& $|\eps^{e,T}_{e\mu}|<0.03$	\\  
		$e^-e^+$ + TEXONO & $|\eps^{e,S}_{e\tau}|<0.28$  & $|\eps^{e,P}_{e\tau}|<0.29$& $|\eps^{e,T}_{e\tau}|<0.07$  	\\  
		$e^-e^+$ + CHARM-II & \multicolumn{2}{c}{$|\eps^{e,X}_{\mu\tau}|<0.25$} & $|\eps^{e,T}_{\mu\tau}|<0.03$	\\  \hline
		CHARM$-e$ & \multicolumn{2}{c}{$|\eps^{q,X}_{ee}|<1.9$}  & $|\eps^{q,T}_{ee}|<0.13$	\\  
		CHARM + CDHS (+ NuTeV) & \multicolumn{2}{c}{$|\eps^{q,X}_{\mu\mu}|<0.15 \, (0.1)$} & $|\eps^{q,T}_{\mu\mu}|<0.01 \, (0.006)$	\\
		CHARM$-e$ + CHARM + CDHS (+ NuTeV) & \multicolumn{2}{c}{$|\eps^{q,X}_{e\mu}|<0.15 \, (0.1)$} & $|\eps^{q,T}_{e\mu}|<0.01 \, (0.006)$	\\
		CHARM$-e$ & \multicolumn{2}{c}{$|\eps^{q,X}_{e\tau}|<1.9$}	& $|\eps^{q,T}_{e\tau}|<0.13$	\\ 
		CHARM + CDHS (+ NuTeV) & \multicolumn{2}{c}{$|\eps^{q,X}_{\mu\tau}|<0.15 \, (0.1)$} & $|\eps^{q,T}_{\mu\tau}|<0.01 \, (0.006)$	\\  \hline
	\end{tabular} 
	\caption{\footnotesize Combined $90\%$ C.L. limits on the different scalar, pseudoscalar, and tensor neutrino interaction parameters, with $X=S,P$. For each suitable parameter, we also show in brackets the corresponding limits including the NuTeV measurements.} \label{tab:combinedLimits}    
\end{table}

\begin{figure}
	\centering
	\includegraphics[width=\linewidth]{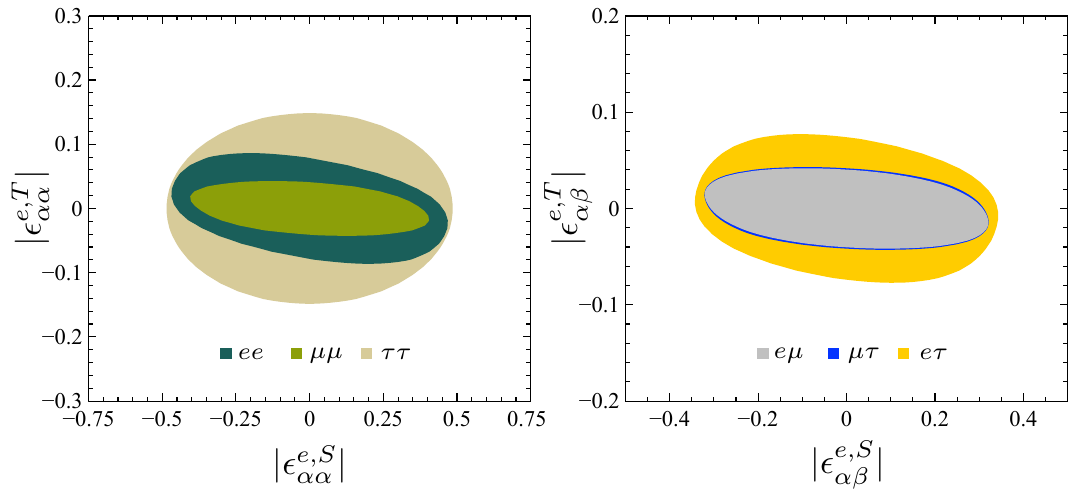}
	\caption{Global constraints on the scalar and tensor neutrino-electron parameters, at $90\%$ C.L. The same combination of
	experiments as in Table~\ref{tab:combinedLimits} was used. The left (right) panel shows the constraints on the diagonal (nondiagonal) parameters.}
	\label{fig:contour_electron}
\end{figure}

\begin{figure}
	\centering
	\includegraphics[width=\linewidth]{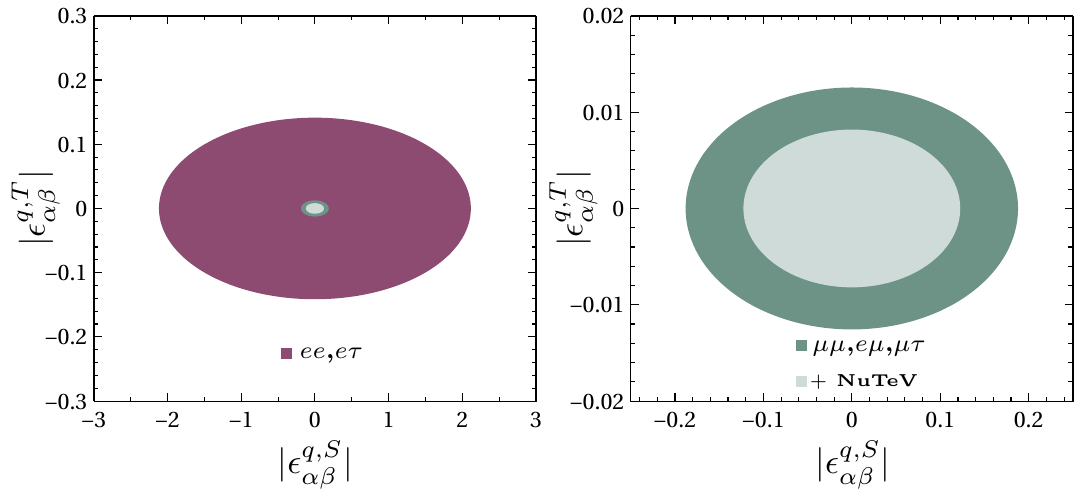}
	\caption{Global constraints on the scalar and tensor neutrino-quark parameters, at $90\%$ C.L. The same combination of
		experiments as in Table~\ref{tab:combinedLimits} was used. In the left panel we show the less stringent constraints
		corresponding to the parameters with flavor index $ee$ and $e\tau$. The other constraints, with and without including
		data from NuTeV, are presented in the right panel. As a reference for the reader, we also include the stronger constraints
		in the left panel.}
	\label{fig:contour_quark}
\end{figure}

For completeness, we include an extra degree of freedom in this second
part of the analysis, carrying out a new calculation where we let two
parameters (of scalar and tensor nature) vary
freely. Figs.~\ref{fig:contour_electron} and~\ref{fig:contour_quark}
show the regions obtained at $90\%$~C.L. for neutrino-electron and
neutrino-quark interactions. As expected, introducing an extra
parameter relaxes the bounds on $\epsilon^{f,S}_{\alpha \beta}$ and
$\epsilon^{f,T}_{\alpha \beta}$. Despite this, constraints coming from
muon-neutrino-quark interactions are still the most restrictive as we
can see in Fig.~\ref{fig:contour_quark}.

 To conclude this section, we make a rough
  estimate of the scale for new physics implied by our constraints.
  As mentioned in the introduction, different models for new physics
  can lead to the couplings discussed here. To cite one example,
  models with leptoquarks, as the one discussed in
  Ref.~\cite{Bischer:2019ttk}, could be an UV completion for GNI.  If
  we consider the four-fermion effective interaction discussed
  here as the low energy limit of a heavy propagator coming from new
  physics, we will conclude that
\begin{equation}
\epsilon\frac{G_{F}}{\sqrt{2}}\simeq\frac{g_X^{2}}{M_X^{2}}\,,
\end{equation}
where $g_X$ and $M_X$ are the coupling and mass of a new physics mediator, respectively. In this case, for neutrino-lepton interactions, our constraints will
imply that $\frac{M_X}{g_X}\approx550$~GeV for the scalar coupling and
$\frac{M_X}{g_X}\approx1.8$~TeV for the tensor case.  For the
neutrino-quark interactions, we will have $\frac{M_X}{g_X}\approx1$~TeV
for the scalar coupling and $\frac{M_X}{g_X}\approx 3.5$~TeV for the tensor
case.  These results are comparable with the
ratios for $M_X/g_X$ presented in~\cite{Bischer:2018zcz} and are of the
same order than in the charged-current sector (see, for example,
Refs.~\cite{Chang:2014iba} and \cite{Gonzalez-Solis:2019owk}).

%

\section{Conclusions}
\label{sec:conclusions}

In this article, we have studied Generalized Neutrino Interactions in
a model-independent way, following an effective field theory approach
with the Standard Model theory at low energies as a framework. Based
on a wide range of experimental results, we have performed a
statistical analysis to find the restrictions on the different GNI
parameters, especially for those of scalar, pseudoscalar, and tensor
nature. For the sake of completeness, for vector and axial cases, we
have quoted previously reported results.

We have studied the individual constraints for each of the nine
experiments commented on previously. Some of these constraints are new,
such as those coming from the electron-positron collision to a
neutrino-antineutrino pair plus photon signal, the CDHS
experiment, and the TEXONO experiment. We have also re-analyzed
the NuTeV anomaly considering the recent results on the systematic
uncertainties and provided new restrictive constraints for the GNI
parameters.  For the scalar and pseudoscalar cases, we have found that
the bounds coming from the electron-positron collision experiments are
of the same order of magnitude as those obtained from CHARM-II and
TEXONO. In constrast,  there is one order of magnitude difference for the
tensor parameters.

We have also performed a robust global analysis for GNI using nine
different neutrino experiments: ALEPH, DELPHI, L3, OPAL, CHARM-II,
TEXONO, CHARM, CDHS, and NuTeV. With this analysis, we have set new
constraints on the GNI parameters that, in some cases, are more
restrictive than previously reported bounds. Being a global analysis, we
have also shown the restrictions for combinations of two parameters at
a time.  We have found that, in general, the interactions with quarks
are more constrained than the ones with electrons.

\section{Acknowledgments}
\label{sec:appendix}

We thank useful discussions with Danny Marfatia. The work of LJF is
supported by a postdoctoral CONACYT grant, CONACYT
CB2017-2018/A1-S-13051 (M\'exico) and DGAPA-PAPIIT
IN107118/IN107621. OGM has been supported by the CONACYT grant
A1-S-23238. The work of J. R. has been granted by his CONACYT
scholarship.

\appendix
\section{Parametrizations for GNI interactions}

There are at least two different parametrizations for GNI
interactions~\cite{Bischer:2019ttk}. For completeness, we show the
relation between them in this appendix. In this work we have used the
so-called epsilon parametrization
\begin{equation}
	\mathcal{L}^{NC}_{eff}=-\frac{G_{F}}{\sqrt{2}}\sum_{j=1}{\epsilon}^{f,j}_{\alpha\beta\gamma\gamma}(\bar{\nu}_{\alpha}\mathcal{O}_{j}\nu_{\beta})(\bar{f}_{\gamma}\mathcal{O}^{'}_{j}f_{\gamma}),
\end{equation}
where we have defined the corresponding operators and couplings in
Table~\ref{table_1}. Note that, although redundant, we have explicitly written the
flavor index in $\epsilon$ and $f$, to have a clear comparison with
the $C$ and $D$ coefficients. In works
\cite{AristizabalSierra:2018eqm,Han:2020pff} they used the CD
parametrization which is given by the following Lagrangian
\begin{equation}
	\mathcal{L}=-\frac{G_{F}}{\sqrt{2}}\sum_{a=S,P,V,A,T}(\bar{\nu}_{\alpha}\Gamma^{a}\nu_{\beta})(\bar{f}_{\gamma}\Gamma^{a}(C^{a}_{\alpha\beta\gamma\gamma}+\bar{D}^{a}_{\alpha\beta\gamma\gamma}i\gamma^{5})f_{\gamma})\,,
\end{equation}
where the five possible independent combinations of Dirac matrices are given by
\begin{equation}
	\Gamma^{a}\in\{1,i\gamma^{5},\gamma^{\mu},\gamma^{\mu}\gamma^{5},\sigma^{\mu\nu}\}\,.
\end{equation}
The relations between the coefficients $C^{a}_{\alpha\beta\gamma\gamma}$ and $D^{a}_{\alpha\beta\gamma\gamma}\equiv\begin{cases}
	\bar{D}^{a}_{\alpha\beta\gamma\gamma} \,(a=S,P,T)\\
	i\bar{D}^{a}_{\alpha\beta\gamma\gamma}\,(a=V,A)\\
\end{cases}$ and the ${\epsilon}^{f,j}_{\alpha\beta\gamma\gamma}$ couplings is given below:

\begin{eqnarray}
	\epsilon^{L}=\frac{1}{4}(C^{V}-D^{V}+C^{A}-D^{A}),\nonumber \\
	\epsilon^{R}=\frac{1}{4}(C^{V}+D^{V}-C^{A}-D^{A}),\nonumber \\
	\epsilon^{S}=\frac{1}{2}(C^{S}+iD^{P}), \\
	-\epsilon^{P}=\frac{1}{2}(C^{P}+iD^{S}), \nonumber\\
	\epsilon^{T}=\frac{1}{4}(C^{T}-iD^{T}). \nonumber
\end{eqnarray}

\bibliography{references}

\end{document}